\documentclass[reprint,superscriptaddress,aps,longbibliography]{revtex4-1}
\usepackage{amssymb}
\usepackage[utf8]{inputenc}
\usepackage{graphicx}
\usepackage{bm,color,amsmath,subfigure}

\begin{document}

\title{Pseudo-spectral Landau-Lifshitz description of magnetization dynamics}

\author{Kyle Rockwell}
\affiliation{Center for Magnetism and Magnetic Nanostructures, University of Colorado Colorado Springs, Colorado Springs, CO 80918, USA}

\author{Joel Hirst}
\affiliation{Materials and Engineering Research Institute, Sheffield Hallam University, Howard Street, Sheffield S1 1WB 22, UK}

\author{Thomas A. Ostler}
\affiliation{Department of Physics and Mathematics, University of Hull, Hull, HU6 7RX, UK}

\author{Ezio Iacocca}
\affiliation{Center for Magnetism and Magnetic Nanostructures, University of Colorado Colorado Springs, Colorado Springs, CO 80918, USA}

\begin{abstract}
Magnetic materials host a wealth of nonlinear dynamics, textures, and topological defects. This is possible due to the competition between strong nonlinearity and dispersion that act at the atomic scale as well as long-range interactions. However, these features are difficult to analytically and numerically study because of the vastly different temporal and spatial scales involved. Here, we present a pseudo-spectral approach for the Landau-Lifshitz equation that invokes energy and momentum conservation embodied in the magnon dispersion relation to accurately describe both atomic and continuum limits. Furthermore, this approach enables analytical study at every scale. We show the applicability of this model in both the continuum and atomic limit by investigating modulational instability and ultrafast evolution of magnetization due to transient grating, respectively, in a 1D ferromagnetic chain with perpendicular magnetic anisotropy. This model provides the possibility of grid-independent multiscale numerical approaches that will enable the description of singularities within a single framework.
\end{abstract}

\maketitle

Nonlinear and far-from-equilibrium dynamics give rise to fascinating physics because of novel phenomena~\cite{Sie2019,Budden2021,Zhou2021} and their potential applications~\cite{Zutic2004,Kirilyuk2010}. Magnetic materials are fundamentally interesting to study from this perspective because their nonlinearity and dispersion gives rise to a wealth of local and nonlocal defects~\cite{Kosevich1990,Donnelly2017,Donnelly2020,Donnelly2021,Turenne2022,Rana2023,DiPietro2023} as well as, e.g., symmetry-dependent domain-wall rearrangement~\cite{Pfau2012,Zusin2022,ZhouHagstrom2022}, ultrafast domain-wall speeds~\cite{Jangid2023}, and stabilization of topological phases~\cite{Iacocca2019,Buettner2021}. 

From a theoretical and numerical point of view, the investigation of such dynamics presents many challenges. For example, solitons typically exhibit a singularity at their core, which requires atomic scale resolution, while its environment is relatively smooth~\cite{Andreas2014,Rana2023}; and in far-from-equilibrium phenomena, both short and long-range effects play a role at picosecond timescales~\cite{Iacocca2019}. There are traditionally two approaches to resolve the dynamics. On the one hand, atomistic spin dynamics (ASD)~\cite{Evans2012,Evans2014,Ostler2011,Barker2019,Kartsev2020,Jakobs2022} take into account a discrete Heisenberg Hamiltonian to resolve the interactions between atomic magnetic moments. On the other hand, micromagnetic simulations~\cite{Brown1963b,Abert2019} are based on the series expansion of the Heisenberg Hamiltonian in a continuum regime~\cite{Krawczyk2012} which can resolve domain structures. Because of the series expansion of the exchange interaction, there is a spatio-temporal transition region that is poorly described by either method, limiting the understanding of several problems of current interest such as the stabilization of topological solitons or domains by ultrafast excitation, and the conditions to nucleate and stabilize three-dimensional solitons~\cite{Rana2023,Balakrishnan2023}.

Here, we introduce a pseudo-spectral Landau-Lifshitz equation (PS-LLE) that closes the gap between ASD and micromagnetic simulations. In contrast to previous multiscale approaches that link ASD and micromagnetic simulations, e.g., Refs.~\cite{Atxitia2010,Andreas2014,Mendez2020,Arjmand2020}, we seek a description that targets the transition region. Our fundamental assumption is that there must be continuity in both energy and momentum, which we enforce by ensuring a proper description of the dispersion relation of magnons, the quanta of angular momentum, across scales. This goal is achieved in Fourier space and it is similar to the ``dispersion engineering'' approach introduced by Whitham in the context of fluid dynamics~\cite{Whitham1974}. In addition, working in Fourier space enables both grid independence as well as analytical study of atomic scale and micrometer scale effects on an equal footing, thus providing a formalism to investigate nonlinear textures.

While the PS-LLE model is general, we focus here on its description and implementation for a one-dimensional (1D) ferromagnetic chain with perpendicular magnetic anisotropy. First, we demonstrate that the magnon dispersion relation is numerically reproduced {and that the computation is grid independent.} Second, we show that modulational instability~\cite{zakharov_modulation_2009} is numerically reproduced, agreeing with analytical expressions obtained through a spin hydrodynamic representation~\cite{Sonin2010,Takei2014,Iacocca2017,Iacocca2019_Rev}. Third, we consider ultrafast transient grating~\cite{Bencivenga2019,Rouxel2021,Ksenzov2021,Foglia2023,Bencivenga2023} as a test case to compare ASD and PS-LLE, demonstrating excellent agreement in the energy evolution and difference in damping between both models. Because of the arbitrary resolution in space, PS-LLE is ideal to investigate topological textures and far-from-equilibrium dynamics. The generality of the model also suggests further refinement by integration with existing micromagnetic approaches, multigrid expansions, and multi-sublattice systems.

The discrepancy in energy and momentum space between ASD and the micromagnetic approach is especially apparent in the magnon dispersion relation. For example, the magnon dispersion relation for 1D ferromagnets with nearest-neighbor interactions can be obtained from the discrete Heisenberg Hamiltonian
\begin{equation}
    \label{eq:omega}
    \omega(k) = \gamma\mu_0M_s\frac{2D}{a^2\hbar}\left(1-\cos{(ka)}\right)\approx\gamma\mu_0M_s\frac{D}{\hbar}k^2,
\end{equation}
where $\gamma$ is the gyromagnetic ratio, $D$ is the exchange stiffness, $\hbar$ is the reduced Planck's constant, and $a$ is the lattice constant of the crystal. Upon first-order Taylor expansion, the dispersion relation reduces to the well-known $k^2$ form obtained in the micromagnetic limit, and also shown in Eq.~\eqref{eq:omega}. In this limit, the energy of magnons at the first Brillouin zone (FBZ) boundary is a factor $\pi^2/4\approx2.46$ larger than in ASD. More dramatically, the group velocity of magnons is substantially different, being exactly zero for the discrete Heisenberg Hamiltonian and $\gamma\mu_0M_sD\pi^2/(\hbar a^2)$ in the micromagnetic limit. This implies that small features, such as topological defects, are difficult to stabilize in micromagnetic simulations.

A similar problem was encountered in the description of shallow water waves, where the Korteveg-de Vries (KdV) equation is a long-wave approximation of the Euler equations~\cite{Carter2018}. In his seminal work, Whitham~\cite{Whitham1974} proposed a mathematical solution whereby the term giving rise to wave dispersion in the KdV equation, $u_{xxx}$, could be modified to exactly reproduce the phase velocity obtained from Euler equations. For this, a kernel $\kappa$ was introduced so that $u_{xxx}\rightarrow\kappa*u_x$, where $*$ represents convolution. Because of the convolution theorem, this method is numerically apt for spectral methods.

In the context of ferromagnets, the magnetization dynamics are described by the Landau-Lifshitz-Gilbert (LLG) equation. To describe the magnon dispersion relation, we introduce the pseudo-spectral Landau-Lifshitz equation (PS-LLE)
\begin{eqnarray}
  \label{eq:pslle}
    \frac{\partial\mathbf{m}}{\partial t} &=& -\mathbf{m}\times\Big[\left(\gamma\mu_oM_\mathrm{eff}\mathbf{h}_\mathrm{l}-\mathcal{F}^{-1}\{\omega(k)~\hat{\mathbf{m}}\}\right)\\
    &&+\alpha\mathbf{m}\times\mathbf{m}\times\left(\gamma\mu_oM_\mathrm{eff}\mathbf{h}_\mathrm{l}-\mathcal{F}^{-1}\{\omega(k)~\hat{\mathbf{m}}\}\right)\Big]\nonumber,
\end{eqnarray}
where $\mathbf{m}$ is the magnetization vector normalized to the saturation magnetization, $M_s$, and $\mathbf{h}_\mathrm{l}$ contains the local fields normalized to the effective magnetization $M_\mathrm{eff}=|H_k-M_s|$, where $H_k$ is the anisotropy field~\cite{Iacocca2017}. The exchange field {and long-range interactions are described in} Fourier space as $\omega(k)\hat{\mathbf{m}}$, where $\omega(k)$ is the desired dispersion relation. In Eq.~\eqref{eq:pslle}, $\mathcal{F}^{-1}\{\cdot\}$ represents the inverse Fourier transform, indicating that the dispersion relation is used as a convolution kernel. This is different than the approach introduced by Whitham~\cite{Whitham1974} which focused on describing the phase velocity rather than the dispersion relation.

The PS-LLE is generally stated in Eq.~\eqref{eq:pslle}, but here we simplify the analysis to the case of a 1D chain subject to a normalized external field $h_o$ and perpendicular magnetic anisotropy (PMA), both normal to the plane, $\mathbf{h}_\mathrm{l}=(h_o+m_z)\hat{z}${, and exchange interaction. Non-local dipole field is not included here to focus on the correct description of the exchange interaction.} It can be readily shown that the magnon dispersion relation $\omega(k)$ is directly recovered from Eq.~\eqref{eq:pslle} by imposing the plane-wave solution $m_z\approx1$ and $m_x,m_y\propto e^{i(kx-\omega t)}$. A detailed derivation is provided in the Supplementary Material~\ref{sec:si_dispersion}~\cite{Supp}.
\begin{figure}[b]
\centering \includegraphics[width=3.4in]{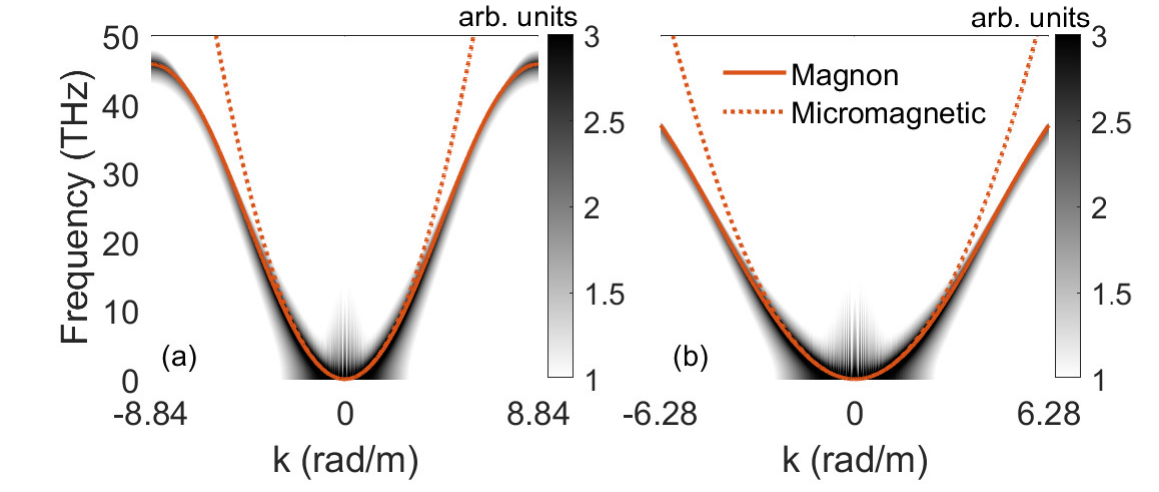}
\caption{ \label{fig:dispersion} Dispersion relation from PS-LLE in a 1D ferromagnetic chain. The cells are discretized to (a) $a=0.355$~nm and (b) $0.5$~nm. The analytical magnon dispersion relation is overlaid with solid red curves and the micromagnetic approximation with red dashed curves. In both cases, the PS-LLE reproduces the magnon dispersion relation.}
\end{figure}

The numerical implementation of PS-LLE also reproduces the correct dispersion relation, shown in Fig.~\ref{fig:dispersion}(a). For this numerical solution, we set parameters for CoFe/Ni, e.g., Ref.~\cite{ZhouHagstrom2022}: $M_s=770$~kA/m, $H_k=1,550$~kA/m, $\lambda_\mathrm{ex}=7.3$~nm, and lattice constant $a\approx0.355$~nm. The 1D ferromagnetic chain has an equilibrium orientation along the $m_z$ component. We introduce a delta-like defect in the center of the chain, so that $m_x=1$ and $m_z=0$ at that point. The evolution of this defect results in outward propagating waves~\cite{El2016} that access all available magnons in the FBZ. We evolve the simulation for $10$~ps with time step of $10$~fs. The numerical result using an atomic cell size is shown in Fig.~\ref{fig:dispersion}(a) with black contrast. We overlay the magnon dispersion relation by a solid red curve and the micromagnetic approximation by a red dashed curve, demonstrating that the PS-LLE method correctly describes the magnon dispersion relation.

To investigate grid independence, we repeat the simulation setting a cell size of $0.5$~nm, which is incommensurate to $a$. The result is shown in Fig.~\ref{fig:dispersion}(b) with black contrast, also displaying excellent agreement with the magnon dispersion relation up to the maximally resolved wavenumber $\pi/0.5$~nm~$=6.28$~rad nm$^{-1}$.

\begin{figure}[t]
\centering \includegraphics[width=3in]{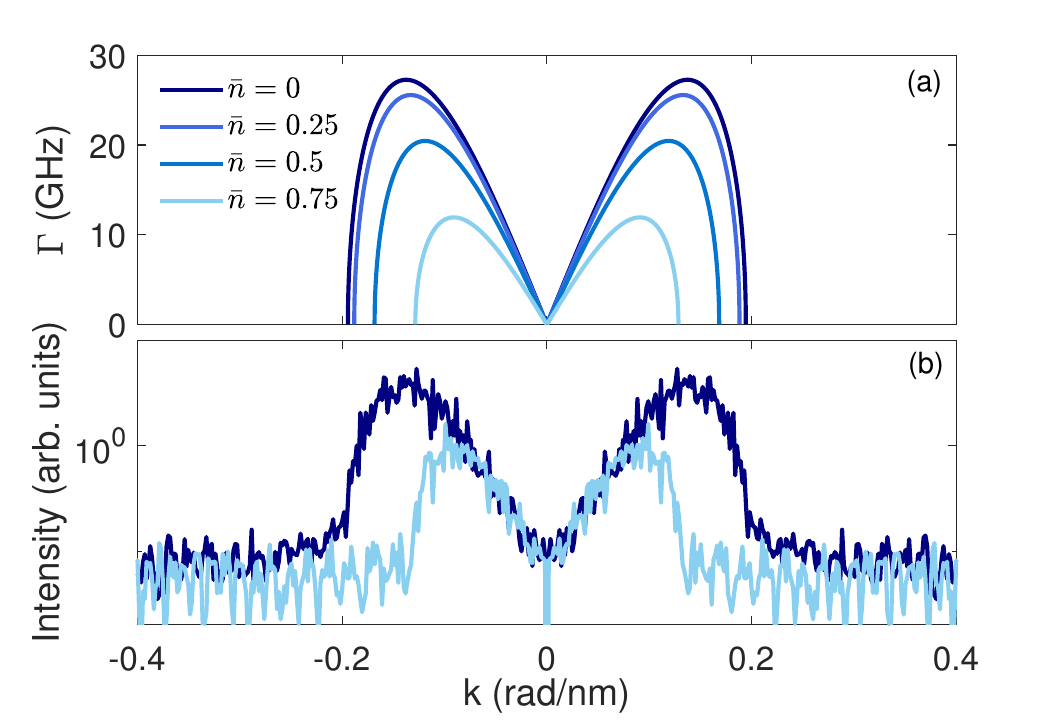}
\caption{ \label{fig:MI} (a) Calculation of the growth rate $\Gamma$ for $\bar{n}=0.25$, $0.5$, $0.75$, and $1$. As the PMA magnet is brought more into the plane, the MI band and maximum growth rate increase. (b) Spectrum of the $m_z$ component after $t=50$~ps obtained by PS-LLE. The simulation is initialized at a constant $m_z$ and perturbed with a small, uniformly distributed noise. The spectrum of 100 instances are averaged to minimize noise in the Fourier transform. Excellent agreement between the spectra and the growth rates are observed.}
\end{figure}

The PS-LLE also enables analytical investigation, insofar as solutions can be obtained in Fourier space. For example, it is possible to apply a spin hydrodynamic formulation~\cite{Sonin2010,Takei2014,Iacocca2017,Iacocca2019_Rev} to the PS-LLE, where we take advantage of the uniaxial symmetry of the local field to introduce a spin density $n=m_z$ and a fluid velocity $u=-\nabla \phi=-\nabla\left[\mathrm{atan}(m_y/m_x)\right]$. This approach simplifies the description of chiral magnetization states and allows one to obtain the dispersion relation of magnons on such states. The full derivation is presented in the Supplementary Material~\ref{sec:si_spinhydro}~\cite{Supp}.

Here, we focus on the long-wave instability of magnons in PMA materials, generally known as modulational instability (MI) for focusing media~\cite{zakharov_modulation_2009}. In its early temporal stage, MI is a linear argument whereby the dispersion of magnons becomes complex under a range of wavevectors. Such wavevectors are typically low and are thus resolved in the micromagnetic limit. Therefore, the PS-LLE must be capable of describing MI. To solve for MI conditions in the spin hydrodynamic formulation, we set $u=0$ (uniform magnetization) and average spin density $\bar{n}$. Large-amplitude perturbations on the background $\bar{n}$ have a dispersion relation $\tilde{\omega}(k)$ given by
\begin{equation}
    \label{eq:disp_ss}
    \tilde{\omega}(k) = \pm\omega(k)\sqrt{1-\frac{\gamma\mu_0M_\mathrm{eff}(1-\bar{n}^2)}{\omega(k)}},
\end{equation}
which is valid for $0\leq\bar{n}<1$. Clearly, $\tilde{\omega}(k)\rightarrow\omega(k)$ when $\bar{n}\rightarrow1$. Solving for a negative argument in the square root, leads to a band of wavenumbers $0\leq k\leq k_\mathrm{max}$, where
\begin{equation}
    \label{eq:kmax}
    k_\mathrm{max} = \frac{1}{a}\mathrm{acos}\left[1-\frac{a^2h_k(1-\bar{n}^2)}{2\lambda_\mathrm{ex}^2}\right].
\end{equation}
For the CoFe/Ni parameters used above, the maximum wavenumber is $0.19$~rad~nm$^{-1}$, which is equivalent to a wavelength of $33$~nm, well within the micromagnetic limit (for reference, half the FBZ would be a $0.71$~nm wavelength). The growth rate, $\Gamma$, is simply given by the imaginary part of Eq.~\eqref{eq:disp_ss} within the MI band. We show the growth rate for various values of $\bar{n}$ in Fig.~\ref{fig:MI}(a). We find growth rates on the order of tens of GHz, equivalent to time constants under $100$~ps. $k_\mathrm{max}$ is reduced as $\bar{n}\rightarrow1$, and MI is completely suppressed in the limit of $\bar{n}=1$. 

Because MI is a linear instability condition, it is possible to directly observe the growth rate from the numerical solution of the PS-LLE. We set a uniform $\bar{n}$ subject to a small, uniformly distributed random perturbation and evolve the simulation for $50$~ps using a cell size equal to the lattice constant $a$. The spectrum of $n$ is shown in Fig.~\ref{fig:MI}(b) for $\bar{n}=0$ and $\bar{n}=0.75$, in excellent agreement with the growth rate and the cut-off wavenumber predicted by Eq.~\eqref{eq:kmax}. For these results, we averaged the spectrum of 100 simulations to reduce noise. These results validate the spin hydrodynamic formulation for the PS-LLE and the correct description of long-wave dynamics by the PS-LLE.

We now investigate a regime in which ASD are required to correctly describe the magnetization dynamics. Ultrafast transient grating~\cite{Bencivenga2019,Rouxel2021,Ksenzov2021,Foglia2023,Bencivenga2023} is an experimental technique where the phase of two femtosecond optical pulses are adjusted to produce a periodic profile on a material. The periodicity can be tuned from $10$~nm to over $100$~nm; this large range of wavevectors lies within the transition regime between ASD and micromagnetic simulations. {ASD simulations are able to describe the dynamics, but at significant computational cost, whereas the micromagnetic description breaks down in this regime.} To test the accuracy of PS-LLE w.r.t. ASD and {the micromagnetic approximation}, we model a quasi-1D ferromagnetic nanowire of dimensions $400$~nm~$\times 1$~nm~$\times 1$~nm subject to periodic boundary conditions on the long axis. Transient grating is modeled as a periodic train of pulses with a Gaussian profile. The number of pulses, $N$, is related to the Gaussian standard deviation, $\sigma$, such that $3N\sigma=400$~nm. To introduce variance on the evolution of the dynamics, the amplitude of the $M_{z}$ component and the phase of the in-plane components are randomized via a uniform random distribution. An example of an initial condition with $N=8$ is shown in the Supplementary Material~\ref{sec:si_transient}~\cite{Supp}.
\begin{figure}[t]
\centering \includegraphics[trim={0in 0.5in 0in 0in}, clip, width=3.5in]{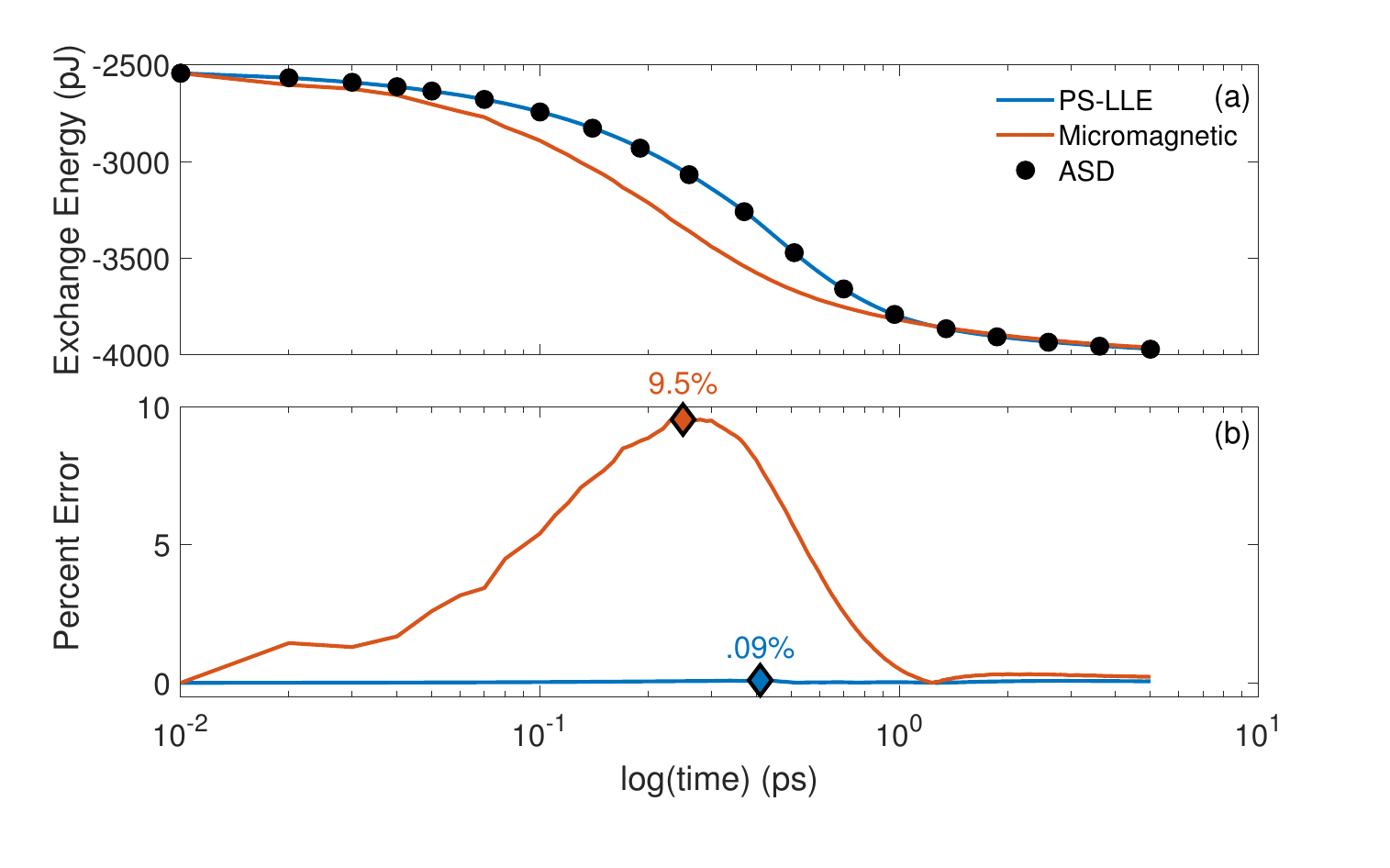}
\caption{ \label{fig:comp} (a) Comparison of exchange energy between PS-LLE in solid blue curve, micromagnetic approximation (LLG) in solid red curve, and ASD in black circles. Excellent agreement with ASD is shown in energy. The associated percent error is shown in (b), with LLG topping at~$9.5\%$ vs. PS-LLE at~$0.09\%$.}
\end{figure}

The initial magnetization state was simulated with ASD and PS-LLE over a timespan of 5 ps. To unambiguously compare with the micromagnetic approximation, we implemented the following pseudo-spectral approach
\begin{eqnarray}
\label{eq:k2}
        \frac{\partial\mathbf{m}}{\partial t} &=&-\mathbf{m}\times\Big[\left(\gamma\mu_oM_\mathrm{eff}\mathbf{h}_\mathrm{l}-\mathcal{F}^{-1}\{k^{2}~\hat{\mathbf{m}}\}\right)\\    &&+\alpha\mathbf{m}\times\mathbf{m}\times\left(\gamma\mu_oM_\mathrm{eff}\mathbf{h}_\mathrm{l}-\mathcal{F}^{-1}\{k^{2}~\hat{\mathbf{m}}\}\right)\Big].\nonumber
\end{eqnarray}
In this case, the kernel is the micromagnetic dispersion relation; this further demonstrates the generalizability of this method. We show that this approach reproduces the micromagnetic magnon dispersion in the Supplementary Material~\ref{sec:si_Laplacian}~\cite{Supp}. 

To compare the methods, we compute the exchange energy as $E=(Aa/2)\sum_i{\mathbf{m_i}\cdot\left(\mathbf{m_{i-1}}+\mathbf{m_{i+1}}\right)}$. The results for an initial state with $N=8$ pulses is shown in Fig.~\ref{fig:comp}(a). The PS-LLE, shown by a solid blue curve, shows a remarkable agreement to ASD, shown with black circles. In contrast, the micromagnetic approximation underestimates the exchange energy under~$\approx1$~ps, stemming for the high-energy of short waves and their concomitant fast damping. The percent energy error is shown in Fig.~\ref{fig:comp}(b), were we note that maximum error between the PS-LLE and ASD $\approx 0.09\%$ in contrast to a maximum of $\approx 9.5\%$ for the micromagnetic approximation. These qualitative trends are maintained for other number of pulses. Results from 1 and 16 pulses are provided in the Supplementary Material~\ref{sec:si_periods}~\cite{Supp}.

\begin{figure}[t]
\centering \includegraphics[width=2.5in]{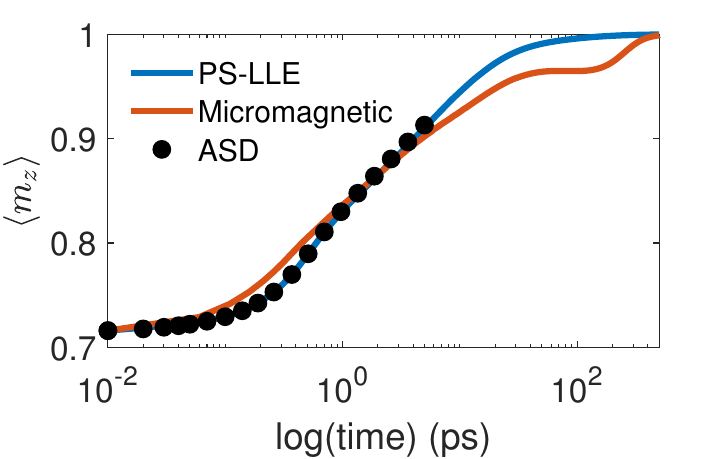}
\caption{ \label{fig:MagEvo} Evolution of the average out-of-plane magnetization, $\langle m_z\rangle$ across the 1D ferromagnetic nanowire, generated from the same dataset as Fig.~\ref{fig:comp}. The results from PS-LLE, micromagnetic approximation, and ASD are shown by solid blue curve, solid red curve, and black circles, respectively. The PS-LLE and its micromagnetic approximation were run for~$500$~ps, until the magnetization reached equilibrium.}
\end{figure}

The energy discrepancy also has consequences in the evolution and recovery of the average magnetization. There is a notable difference between the group velocities obtain from the magnon dispersion and its micromagnetic approximation. In the former, $v_g=2\gamma \mu_{0} M_{s}(D/\hbar a)\sin(ka)$ while the micromagnetic approximation leads to $v_g\approx 2\gamma \mu_{0} M_{s} (D/\hbar)k$. As the micromagnetic group velocity is linear, it quickly diverges from the true group velocity, and thus the energy across the system dissipates faster as the wavenumber increases. This issue is apparent from the evolution of the spatially averaged magnetization along the $z$-axis, $\langle m_z\rangle$, shown in Fig.~\ref{fig:MagEvo}. The ASD results shown by black circles are obtained from the same simulation presented in Fig.~\ref{fig:comp}. The evolution using a micromagnetic approximation, shown by a red curve, over-estimates $\langle m_z\rangle$ from from~$\approx0.5-1$~ps and later under-estimates $\langle m_z\rangle$ until the end of the simulation at $500$~ps. This is a consequence of the disparate group velocities between ASD and the micromagnetic approximation. In contrast, the PS-LLE shown by a solid blue curve follows the ASD evolution accurately and then smoothly stabilizes to full saturation, converging to the micromagnetic description.

We note that the same discrepancy in group velocities is observed in the context of MI. Examples of such an evolution as shown in the Supplementary Material~\ref{sec:si_MI}~\cite{Supp}.

In summary, we have introduced the PS-LLE as a continuum model that captures the relevant physics of atomic scale magnetism by dispersion engineering. {This approach relies on the knowledge of the dispersion relation of magnons and so it can be generalized to higher dimensions, other magnetic orders, and arbitrary exchange integral between spins in a multilattice material.} 
In the context of ultrafast magnetism, excellent agreement with ASD suggests that the PS-LLE can resolve the multiscale evolution of the magnetization without incurring in the violation of energy and momentum conservation. It is important to note that, as a continuum model, the PS-LLE cannot capture the discrete interactions between atoms so it is not expected to be accurate when modeling light-matter interaction via three-temperature's model. However, we shown in the Supplementary Material~\ref{sec:si_TMM}~\cite{Supp} that the PS-LLE closely follows the ASD evolution from initial conditions obtained by modeling transient grating with a laser profile once a quasi-equilibrium temperature has been reached. The PS-LLE will be valuable to provide both analytical and numerical insights into the short time evolution of far-from-equilibrium magnetization, including the onset and evolution of topological defects~\cite{Rana2023,Balakrishnan2023}.

\section*{Acknowledgments}\
This work was supported by the U.S. Department of Energy, Office of Basic Energy Sciences under Award Number DE-SC-0024339. The ASD simulations were completed using resources provided by the Baskerville Tier 2 HPC service. Baskerville was funded by the EPSRC and UKRI through the World Class Labs scheme (EP/T022221/1) and the Digital Research Infrastructure programme (EP/W032244/1) and is operated by Advanced Research Computing at the University of Birmingham.

\onecolumngrid
\clearpage

\Large{\textbf{Supplementary material: Pseudo-spectral Landau-Lifshitz description of magnetization dynamics}}
\\

\large{Kyle Rockwell et al.}
\newpage

\setcounter{section}{0}
\renewcommand{\thesection}{SI \arabic{section}}

\setcounter{figure}{0}
\renewcommand{\thefigure}{SI \arabic{figure}}

\setcounter{table}{0}
\renewcommand{\thetable}{SI \Roman{table}}

\setcounter{equation}{0}
\renewcommand{\theequation}{SI \arabic{equation}}

\normalsize
\section{Derivation of dispersion relation for a 1D magnetized ferromagnetic chain}
\label{sec:si_dispersion}

We start from the conservative PS-LLE, i.e., when $\alpha=0$
\begin{equation}
  \label{eq:si_der1}
    \frac{\partial\mathbf{m}}{\partial t} = -\mathbf{m}\times\left(\gamma\mu_oM_s\mathbf{h}_\mathrm{l}-\mathcal{F}^{-1}\{\omega(k)\hat{\mathbf{m}}\}\right).
\end{equation}

Consider a 1D ferromagnetic chain with in-plane anisotropy that is magnetized in the $z$ direction by an external field. Therefore, $\mathbf{h}_\mathrm{l}=(h_0+m_z)\hat{z}$. We can now perturb the magnetization with a small-amplitude wave with frequency $\Omega$ such that $m_z\approx1$, $m_x=Ae^{i(kx-\Omega(k)t)}$, and $m_y=Be^{i(kx-\Omega(k)t)}$, with $A,B\ll 1$, and separate Eq.~\eqref{eq:si_der1} into the magnetization components
\begin{subequations}
\label{eq:si_der2}
\begin{eqnarray}
    \label{eq:si_der21}
    \partial_tm_x &=& -\left[m_y(\gamma\mu_0M_sh_0+\gamma\mu_0M_sm_z-\mathcal{F}^{-1}\{\omega(k)\hat{m}_z\})+m_z\mathcal{F}^{-1}\{\omega(k)\hat{m}_y\}\right],\\
    \label{eq:si_der22}
    \partial_tm_y &=& -\left[-m_z\mathcal{F}^{-1}\{\omega(k)\hat{m}_x\}-m_x(\gamma\mu_0M_Sh_0+\gamma\mu_0M_sm_z-\mathcal{F}^{-1}\{\omega(k)\hat{m}_z\})\right],\\
   \label{eq:si_der23}
   \partial_t m_z &=& -\left[-m_x\mathcal{F}^{-1}\{\omega(k)\hat{m}_y\}+m_y\mathcal{F}^{-1}\{\omega(k)\hat{m}_x\}\right].
\end{eqnarray}
\end{subequations}

Equation~\eqref{eq:si_der23} is of order $AB$ so that it is a next order expansion and can be neglected. This ensures that $\partial_tm_z\approx0$, which is our initial assumption. We now note that $m_z=1$ means that $\hat{m}_z=\delta{k}$ so that the terms $\mathcal{F}^{-1}\{\omega(k)\hat{m}_z\}$ vanish since $\omega(k)=0$ for $k=0$. Performing a Fourier transform of Eqs.~\eqref{eq:si_der21} and \eqref{eq:si_der22} with the above consideration yields
\begin{subequations}
\label{eq:si_der3}
\begin{eqnarray}
    \label{eq:si_der31}
    -i\Omega\hat{m}_x &=& -\hat{m}_y\left(\gamma\mu_0M_sh_0+\gamma\mu_0M_s+\omega(k)\right),\\
    \label{eq:si_der32}
    -i\Omega\hat{m}_y &=& -\hat{m}_x\left(-\omega(k)-\gamma\mu_0M_Sh_0+\gamma\mu_0M_s\right).
\end{eqnarray}
\end{subequations}

This is a simple $2\times2$ eigenvalue problem that can be written in matrix form
\begin{equation}
    \label{eq:si_der4}
    \begin{bmatrix}
    i\Omega & -\left(\gamma\mu_0M_sh_0+\gamma\mu_0M_s+\omega(k)\right)\\
    \left(\gamma\mu_0M_sh_0+\gamma\mu_0M_s+\omega(k)\right) & i\Omega
    \end{bmatrix}
    \begin{bmatrix}
    \hat{m}_x \\ \hat{m}_y
    \end{bmatrix}=0.
\end{equation}

Looking or a family of solutions, we solve for the determinant of the matrix and directly obtain
\begin{equation}
    \label{eq:si_der5}
    \Omega = \pm \left[\gamma\mu_0M_s(h_0+M_s)+\omega(k)\right],
\end{equation}
which is the dispersion relation of magnons, shifted at $k=0$ by the local field.

\section{Derivation of pseudo-spectral spin hydrodynamic formulation}
\label{sec:si_spinhydro}

In the micromagnetic limit, a dispersive hydrodynamic formulation of magnetization dynamics, or spin hydrodynamics, was obtained exactly~\cite{Iacocca2017} by performing the transformation
\begin{equation}
    \label{eq:si_sh1}
    n=m_z,\quad\mathbf{u}=-\nabla\phi=-\nabla\mathrm{arctan}\left[\frac{m_y}{m_x}\right],
\end{equation}
where $n$ is interpreted as a spin density and $\mathbf{u}$ as a fluid velocity. The spin density represents a basis of angular momentum, and the $z$ component is chosen due to the U(1) symmetry of the system~\cite{Takei2014}. In contrast to ``regular'' fluids, the spin density is signed, i.e.m $-1\leq n\leq 1$. The fluid velocity is an interpretation of the phase gradient, where $\phi$ is the in-plane magnetization phase. From definition, the fluid interpretation is irrotational and $\phi$ is only defined if $|n|<1$.

An interesting aspect of spin hydrodynamics is that the fluid velocity arises from magnetization textures, not dynamics per se. In the case of a 1D texture, it is akin to the chirality of the texture. For this reason, the dispersion relation has a non-reciprocal term. For the sake of comparison with the PS-LLE model below, we write here the 1D limit of the spin hydrodynamic equations ($\nabla\rightarrow\partial_x$):
\begin{subequations}
\label{eq:si_sh2}
\begin{eqnarray}
    \label{eq:si_sh21}
    \partial_tn&=&\partial_x\left[(1-n^2)u\right],\\
    \label{eq:si_sh22}
    \partial u &=& -\partial_x\left[(h_k+u^2)n\right]-\partial_x\left[\frac{\partial_{xx}n}{1-n^2}+\frac{n|\partial_xn|^2}{(1-n^2)^2}\right],
\end{eqnarray}
\end{subequations}
where time is scaled to $\gamma\mu_0M_s$ and space is scaled to $\lambda_\mathrm{ex}^{-1}$.

Equations~\ref{eq:si_sh2} have a simple solution termed uniform hydrodynamic states (UHS), where both $n$ and $u$ are constant: $\bar{n}$ and $\bar{u}$. It is the possible to find the dispersion relation of waves on top a UHS. Setting plane waves, $n=\bar{n}+\tilde{n}$ and $u=\bar{u}+\tilde{u}$, where $\tilde{n}=Ae^{i(kx-\omega t}$ and $\tilde{u}=Be^{i(kx-\omega t}$. Inserting into Eqs.~\ref{eq:si_sh2}, we find
\begin{subequations}
\label{eq:si_sh3}
\begin{eqnarray}
    \label{eq:si_sh31}
    \partial_t\tilde{n}&=&(1-\bar{n}^2)\tilde{u}-2\bar{n}\bar{u}\partial_x\tilde{n},\\
    \label{eq:si_sh32}
    \partial\tilde{u} &=& -(h_k+\bar{u}^2)\partial_x\tilde{n}-2\bar{u}\bar{n}\partial_x\tilde{u}-\partial_x\left[\frac{\partial_{xx}\tilde{n}}{1-\bar{n}^2}\right].
\end{eqnarray}
\end{subequations}

Solving the time and spatial derivatives leads to a system of algebraic equations, from which the dispersion relation can be obtained
\begin{equation}
    \label{eq:si_sh4}
    \omega_n = 2\bar{n}\bar{u}k\pm k\sqrt{(1-\bar{n}^2)(h_k+\bar{u}^2)+k^2}
\end{equation}

For the PS-LLE, an exact transformation is not possible because of the convolution with the kernel. This means that certain vectors identities used in the micromagnetic limit are not possible to achieve. However, we can obtain an exact set of equations for the waves on top of a UHS. Because the kernel is convoluted with $\mathbf{m}$, it is convenient to work with the magnetization components expressed in fluid variables,
\begin{equation}
    \label{eq:si_sh5}
    m_x = \sqrt{1-n^2}\cos{(\bar{u}x)},\quad m_y = \sqrt{1-n^2}\sin{(\bar{u}x)},\quad m_z=n.
\end{equation}

Inserting into the PS-LLE, rewriting into fluid form, and performing Fourier transform leads to the relevant coupled equations. We also work in the general case where $\mathcal{F}\{\kappa\}=\omega(k)$. The main steps of the derivation are laid out below.

\underline{The spin density} is given directly by the time dependence of $m_z$
\begin{equation}
    \label{eq:si_sh6}
    \partial_t m_z = -\left[-m_x\mathcal{F}^{-1}\{\omega(k)\hat{m}_y\}+m_y\mathcal{F}^{-1}\{\omega(k)\hat{m}_x\}\right].
\end{equation}

Because $m_x$ and $m_y$ are products of the fluid variables, their Fourier transform becomes a convolution. We assume small waves, so that $1-n^2\approx1-\bar{n}^2$. Using convolution properties and some algebra, we obtain
\begin{equation}
    \label{eq:si_sh7}
    \partial_t\hat{\tilde{n}} = -\bar{n}\mathcal{F}\{\kappa*\sin{(\bar{u}x)}\}[1+\delta(k)]\hat{\tilde{n}}+i\frac{(1-\bar{n}^2)}{k}\mathcal{F}\{\kappa*\cos{(\bar{u}x)}\}[1-\delta(k)]\hat{\tilde{u}}.
\end{equation}

\underline{The phase} is obtained from the definition
\begin{equation}
    \label{eq:si_sh8}
    \partial_t\phi = \frac{m_x\partial_tm_y-m_y\partial_tm_x}{1-m_z^2},
\end{equation}
and where
\begin{subequations}
\label{eq:si_sh9}
\begin{eqnarray}
    \label{eq:si_sh91}
    \partial_tm_x &=& -\left[m_y(\gamma\mu_0H_0+\gamma\mu_0H_km_z-\mathcal{F}^{-1}\{\omega(k)\hat{m}_z\})+m_z\mathcal{F}^{-1}\{\omega(k)\hat{m}_y\}\right]\\
    \label{eq:si_sh92}
    \partial_tm_y &=& -\left[-m_z\mathcal{F}^{-1}\{\omega(k)\hat{m}_x\}-m_x(\gamma\mu_0H_0+\gamma\mu_0H_km_z-\mathcal{F}^{-1}\{\omega(k)\hat{m}_z\})\right].
\end{eqnarray}
\end{subequations}

Inserting Eqs.~\eqref{eq:si_sh9} into Eq.~\eqref{eq:si_sh8} and performing the convolutions ultimately leads to
\begin{equation}
    \label{eq:si_sh10}
    \partial_t\hat{\tilde{u}} = -ik\left[\gamma\mu_0Hk+\omega(k)-\mathcal{F}\{\kappa*\cos{(\bar{u}x)}\}\left(\frac{\bar{n}^2-\delta(k)}{1-\bar{n}^2}\right)\right]\hat{\tilde{n}}-\bar{n}\mathcal{F}\{\kappa*\sin{(\bar{u}x)}\}[1-\delta(k)]\hat{\tilde{u}}.
\end{equation}

\textbf{Equations~\eqref{eq:si_sh7} and \eqref{eq:si_sh10} are the main results.}

For the particular case $\omega(k)=2\gamma\mu_0M_s\Lambda^2(1-\cos{(ka)})$, with $\Lambda=\lambda_\mathrm{ex}/a$, the equations reduce to
\begin{subequations}
\label{eq:si_sh11}
\begin{eqnarray}
    \label{eq:si_sh111}
    -i\omega_n\hat{\tilde{n}} &=& -2\Lambda^2\bar{n}\sin{(ka)}\sin{(\bar{u}a)}\hat{\tilde{n}}+2i\Lambda^2\frac{(1-\bar{n}^2)}{k}\cos{(\bar{u}a)}(1-\cos{(ka)})\hat{\tilde{u}}.\\
    \label{eq:si_sh112}
    -i\omega_n\hat{\tilde{u}} &=& \left[-ik\left(h_k+2\Lambda^2\cos{(ka)}\right)+2i\Lambda^2k\cos{(\bar{u}a)}\left(\frac{1-\bar{n}^2\cos{(ka)}}{1\bar{n}^2}\right)\right]\hat{\tilde{n}}\nonumber\\&&-2i\bar{n}\Lambda^2\sin{(\bar{u}a)}\sin{(ka)}\hat{\tilde{u}}.
\end{eqnarray}
\end{subequations}

Solving this algebraic set of equations leads to the dispersion
\begin{equation}
    \label{eq:si_sh12}
    \omega_n = 2\bar{n}\Lambda^2\sin{(\bar{u}a)}\sin{(ka)}\pm\sqrt{\mathcal{A}\mathcal{B}},
\end{equation}
where
\begin{eqnarray}
   \label{eq:si_sh13}
   \mathcal{A}&=&2\Lambda^2\frac{(1-\bar{n}^2)}{k}\cos{(\bar{u}a)}(1-\cos{(ka)}),\\
   \label{eq:si_sh14}
   \mathcal{B}&=&\left[-k\left(h_k+2\Lambda^2\cos{(ka)}\right)+2\Lambda^2k\cos{(\bar{u}a)}\left(\frac{1-\bar{n}^2\cos{(ka)}}{1-\bar{n}^2}\right)\right].
\end{eqnarray}
and $\omega_n$ is expressed in units of $\gamma\mu_0M_s$. The first term in Eq.~\eqref{eq:si_sh12} is non-reciprocal and can exhibit a wavenumber-dependent sign change. This is in contrast with the spin hydrodynamic formulation from the micromagnetic approximation where a given fluid velocity sign determined the non-reciprocal wave propagation.

In the micromagnetic limit, we see that $\sin{(ka)}\approx ka$, $\sin{(\bar{u}a)}\approx \bar{u}a$, $\cos{(ka)}\approx1-(ka)^2/2$, and $\cos{(\bar{u}a)}\approx1-(\bar{u}a)^2/2$. Therefore, we find that
\begin{eqnarray}
   \label{eq:si_sh15}
   \mathcal{A}&\approx&2\lambda_\mathrm{ex}^2(1-\bar{n}^2)k,\\
   \label{eq:si_sh16}
   \mathcal{B}&\approx&-k\left(h_k+\lambda_\mathrm{ex}^2\bar{u}^2\right)+\frac{\lambda_\mathrm{ex}^2k^3}{1-\bar{n}^2},
\end{eqnarray}
and
\begin{equation}
    \label{eq:si_sh17}
    \omega_n \approx 2\bar{n}\lambda_\mathrm{ex}^2\bar{u}k\pm (\lambda_\mathrm{ex}k)\sqrt{(1-\bar{n}^2)\left(h_k+\lambda_\mathrm{ex}^2\bar{u}^2\right)+\left(\lambda_\mathrm{ex}k\right)^2}.
\end{equation}
where higher-order corrections were neglected. Clearly, this equation agrees with Eq.~\eqref{eq:si_sh4} upon further space scaling by the exchange length.

\section{Model of transient grating initial condition}
\label{sec:si_transient}

\begin{figure}[!h]
\centering 
\includegraphics[trim={1in, 0, 1in, 0}, clip, width=.8\linewidth]{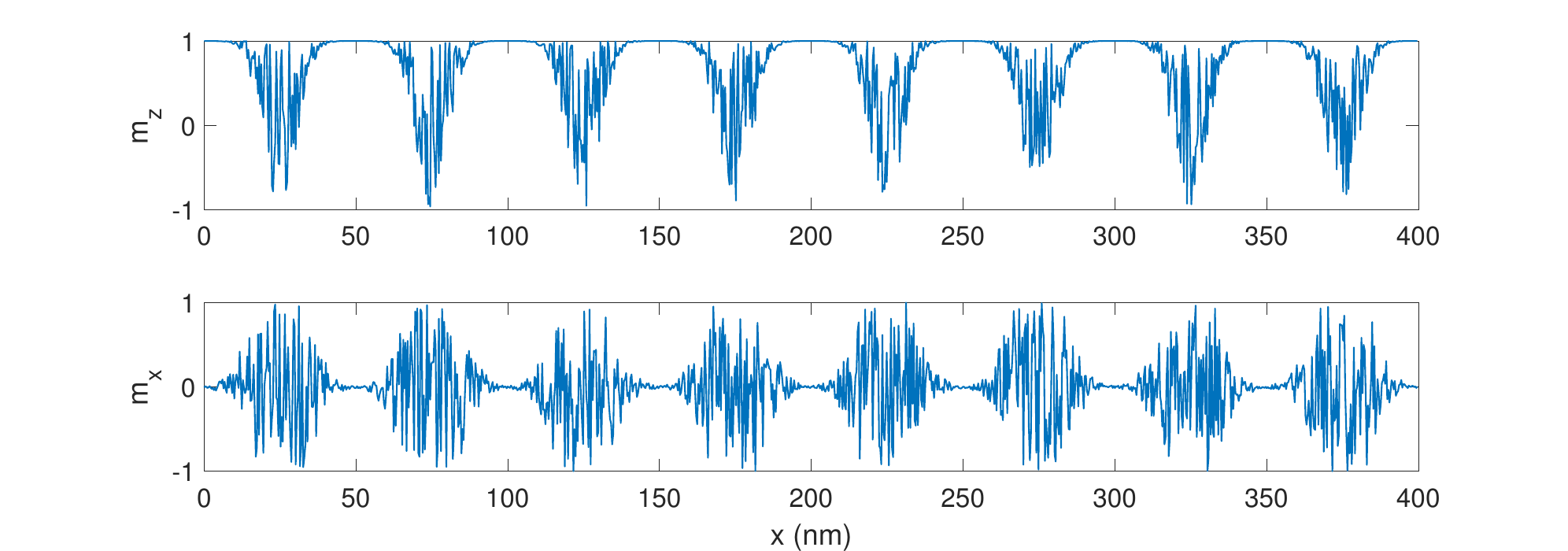}
\caption{Initial magnetization for a train of 8 Gaussian pulses in the 1D ferromagnetic chain. (a) $m_z$ component and (b) $m_x$ component.}
\end{figure}

\section{Atomistic spin dynamics methods}
\label{sec:si_gratingASD}

Atomistic simulations were performed using the Landau-Lifshitz-Gilbert (LLG) equation of motion:
\begin{equation}
\frac{\partial \mathbf{S}_i}{\partial t}=-\frac{\gamma_i}{\left(1+\lambda_i^2\right) \mu_i}\left(\mathbf{S}_i \times \mathbf{H}_i+\lambda_i \mathbf{S}_i \times \mathbf{S}_i \times \mathbf{H}_i\right)
\end{equation}
where $\mathbf{S}_i$ is a normalized unit vector of the spin at site $i$, $\lambda_i$ is the effective damping parameter, $\gamma = 1.76 \times 10^{-11}$ T$^{-1}$s$^{-1}$ is the gyromagnetic ratio, $\mu_i$ is the magnetic moment, and $\mathbf{H}_i$ is the effective field acting on the spin at site $i$. The effective field is given by the equation:
\begin{equation}
    \mathbf{H}_i=\boldsymbol{\zeta}_i(t)-\frac{\partial \mathcal{H}}{\partial \mathbf{S}_i}
\end{equation}
where $\boldsymbol{\zeta}_i(t)$ describes the coupling to the thermal bath and  $\mathcal{H}$ is the Heisenberg Hamiltonian given by:
\begin{equation}
\mathcal{H}=\sum_{i \neq j} J_{i j} \mathbf{S}_i \cdot \mathbf{S}_j-\sum_i d_z S_{i, z}^2
\end{equation}
The CoFe/Ni multilayer structure was treated as a 1D spin chain with periodic boundary conditions using an ``average'' moment model derived from material parameters presented in the main text. In the atomistic model, we use parameters $\mu_s = 3.71 \mu_B$, $d_z = 3.35 \times 10^{-23}$ J, $J_{ij} = 1.41 \times 10^{-20}$ J and $\lambda = 0.01$. These have been derived from the micromagnetic material parameters presented in the main text. The LLG equation is integrated using the Heun method \cite{Nowak2007}.

\section{Micromagnetic validation of the PS-LLE model}
\label{sec:si_Laplacian}

\begin{figure}[!h]
\centering 
\includegraphics[width=4in]{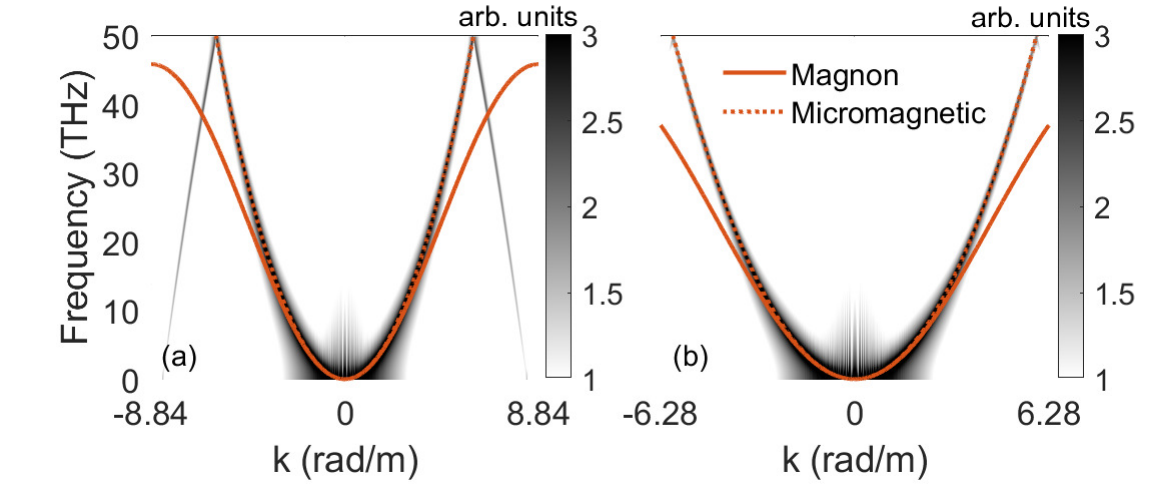}
\caption{Validation the pseudo-spectral implementation of the micromagnetic limit, Eq.~\eqref{eq:k2}, in a 1D ferromagnetic chain using CoFe/Ni parameters. The cells are discretized to (a) $a=0.355$~nm and (b) $0.5$~nm. The analytical magnon dispersion relation is overlaid with solid red curves and the micromagnetic approximation with red dashed curves. In contrast to Fig.~\ref{fig:dispersion} in the main text, this implementation exactly agrees with the micromagnetic dispersion. The reflected branches in (a) occur because of the maximum resolved frequency of $50$~THz is reached.}
\end{figure}

\section{Transient grating for other periods}
\label{sec:si_periods}
\begin{figure}[!h]
\centering 
\subfigure{
\includegraphics[trim={0.2in, 1.012in, 1in, 0}, clip, width=.5\linewidth]{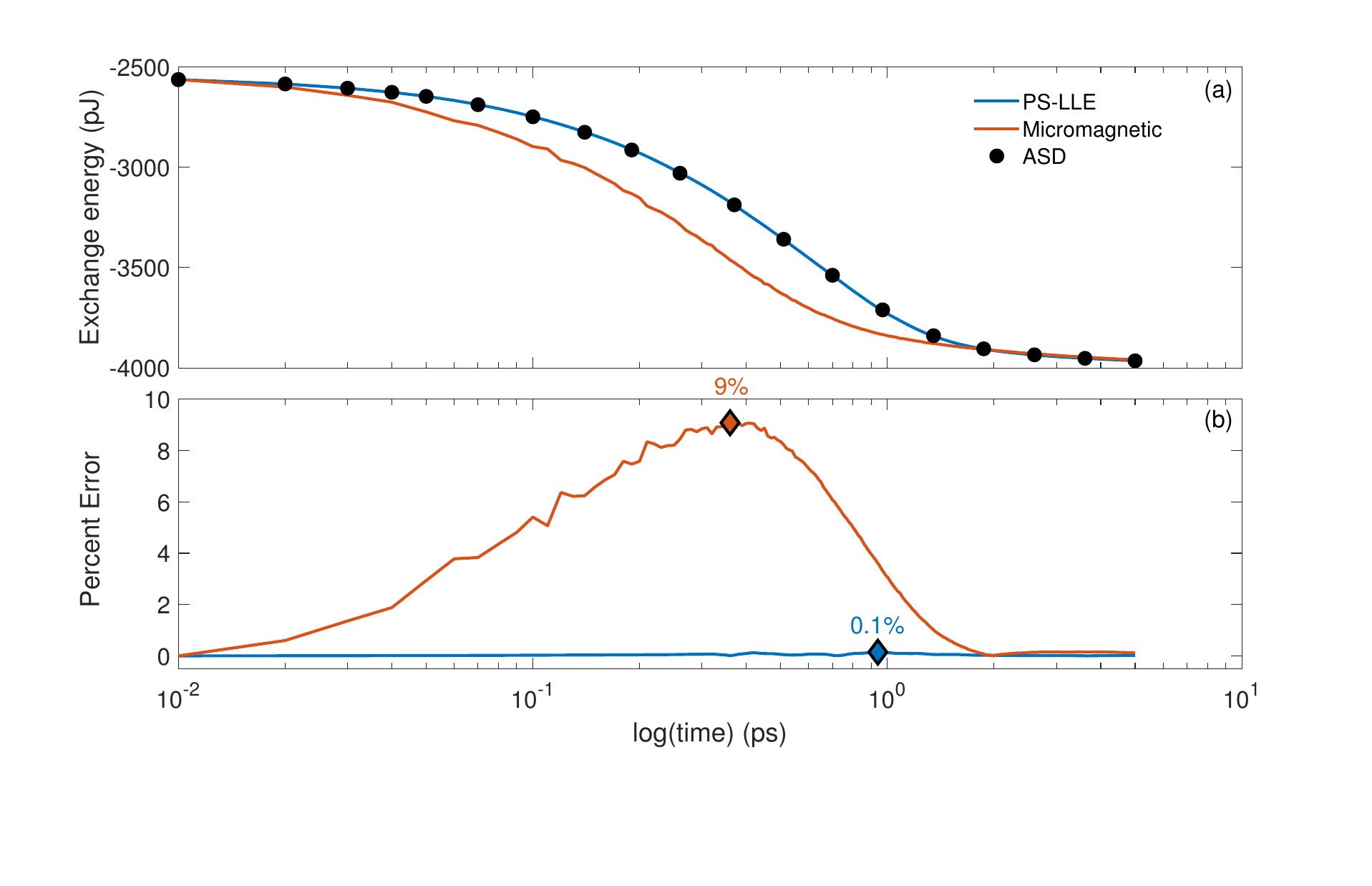}
}
~
\subfigure{
\includegraphics[width=.3\linewidth]{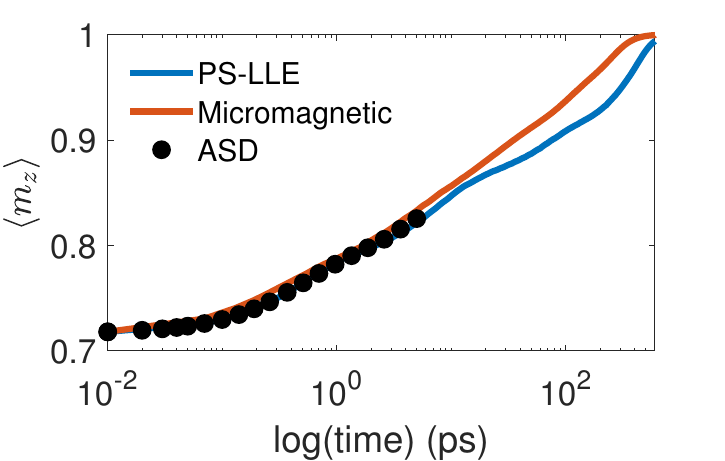}
}
\caption{\label{fig:comp1} On the left is exchange energy and error from the 1 pulse dataset. As seen before in Fig.~\ref{fig:comp}, (a) PS-LLE (blue curve) shows an exceptional agreement with ASD (black circles) compared to the micromagnetic LLG (red curve), where in (b), the micromagnetic LLG produces an error of~$9\%$ v.s PS-LLE at~$0.1\%$. On the right is the magnetization~$\langle m_{z}\rangle$~ over the $500$~ps simulation.}
\end{figure}

\begin{figure}[!h]
\centering 
\subfigure{
\includegraphics[trim={0.2in, .705in, 0.5in, 0}, clip,width=.5\linewidth]{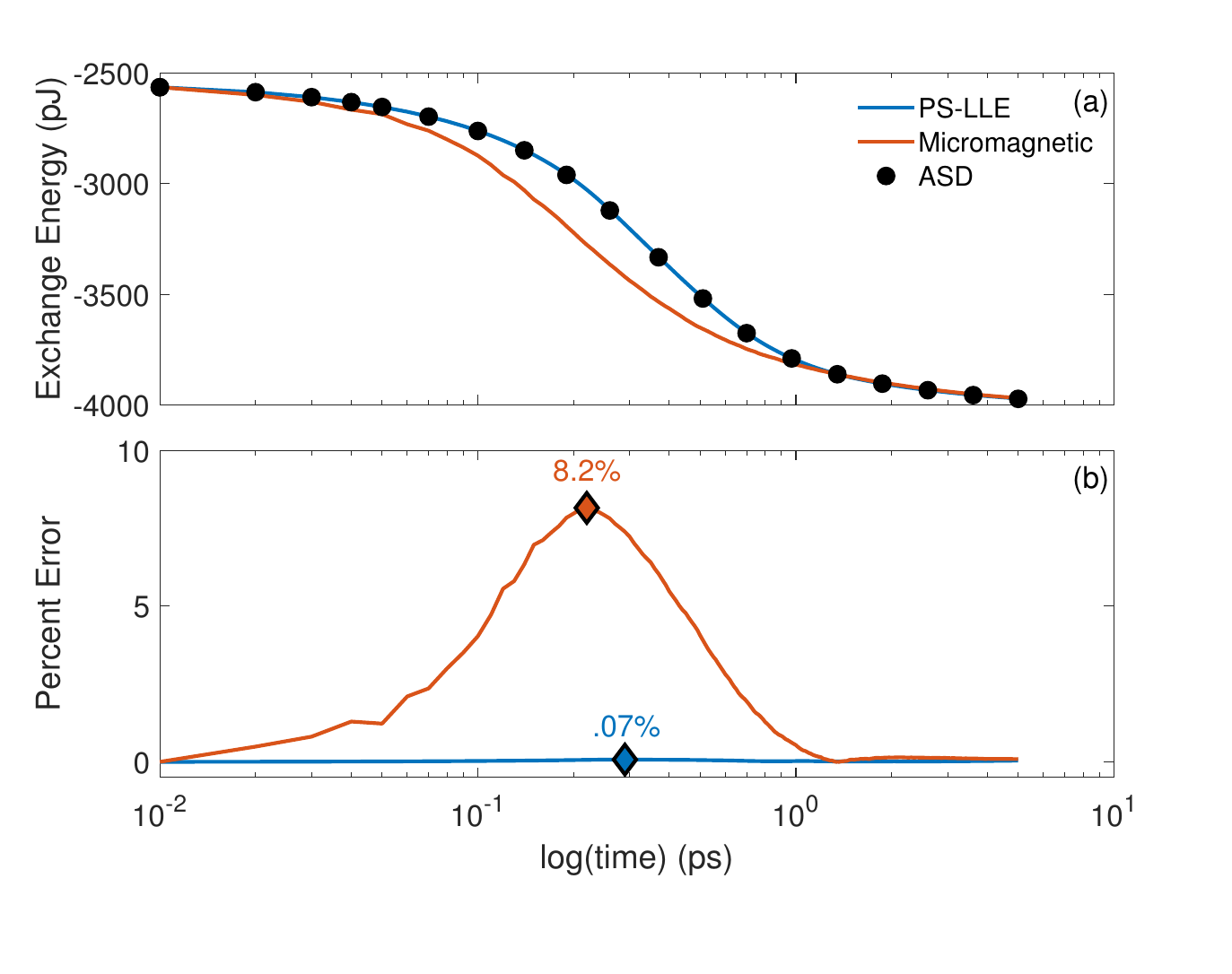}
}
\subfigure{

\includegraphics[width=.3\linewidth]{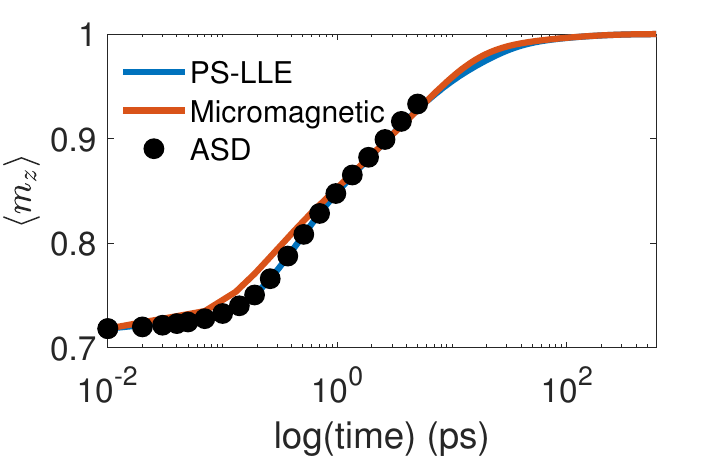}

}
\caption{\label{fig:comp16} On the left is exchange energy and error from the 16 pulse dataset. As seen before in Fig.~\ref{fig:comp}, (a) PS-LLE (blue curve) shows an exceptional agreement with ASD (black circles) compared to the micromagnetic LLG (red curve). A comparison of error is shown in (b), where LLG produces~$8.2\%$ v.s PS-LLE at~$0.7\%$. On the right is the magnetization~$\langle m_{z}\rangle$~evolved for $500$~ps.}
\end{figure}

The resulting energy and magnetization from the single pulse~$5$~ps simulation is shown in Fig.~\ref{fig:comp1}. The energy is plotted with PS-LLE shown in solid blue curve, the micromagnetic approximation in solid red curve, and ASD shown via solid black circles. As seen for the case of 8 pulses in the main text [Fig.~\ref{fig:comp}], PS-LLE outperforms with an error of only~$0.1\%$~compared to~$9\%$~from the micromagnetic LLG. Over the $500$~ps simulation, the micromagnetic LLG reaches equilibrium faster than PS-LLE. This further expresses the discrepancy in the group velocity, seen before in Fig.~\ref{fig:MagEvo}. This can also be noticed in the rate at which the energy decays in Fig.~\ref{fig:comp1}(a). Despite the energy only being computed over~$5$~ps, the micromagnetic approximation consistently predicts less energy and more damping than PS-LLE and ASD. 

Furthermore, the energy for the 16 pulse simulation is shown in Fig.~\ref{fig:comp16}, with the same time spans as in previous cases. As seen previously, PS-LLE produces an error of~$0.07\%$~vs. ~$8.2\%$~from the micromagnetic approximation. With 16 pulses, the group velocity of the magnons is far greater; this should in turn lead to faster damping in energy, causing the magnetization to reach equilibrium sooner. This prediction is verified by Fig.~\ref{fig:comp16}, where the difference in the exchange energy is less than in previous cases and agrees with the micromagnetic approximation earlier than before. Furthermore, the magnetization~$\langle m_{z}\rangle$~predicted by PS-LLE has a much closer adherence to that of the micromagnetic approximation, especially later in the simulation. This further replicates the expected result of faster stabilization when the group velocity of magnons is higher.

\section{Transient grating results: Modulational instability}
\label{sec:si_MI}
\begin{figure}[!h]
\centering 
\includegraphics[width=2.5in]{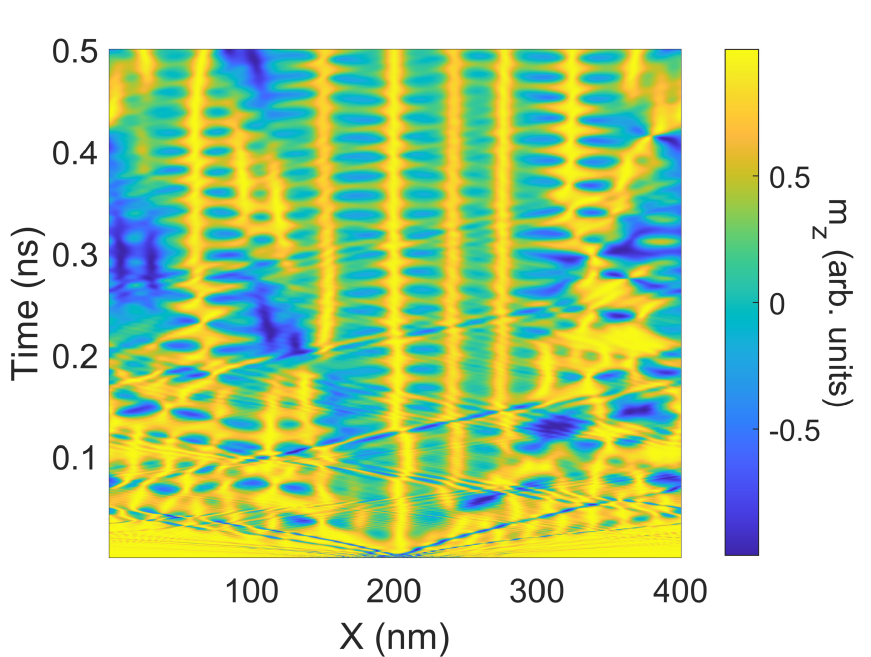}
\includegraphics[width=2.5in]{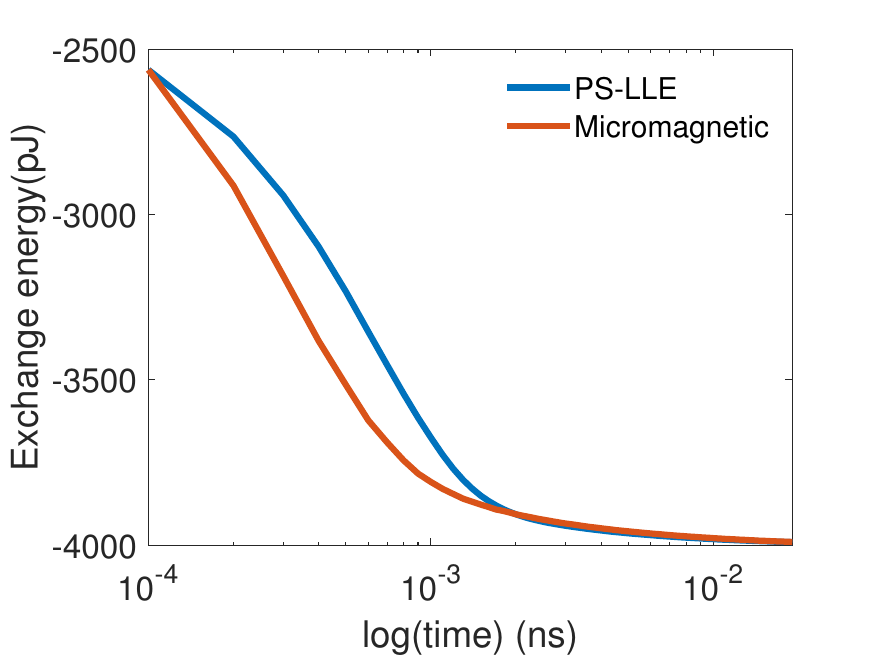}
\includegraphics[width=2.5in]{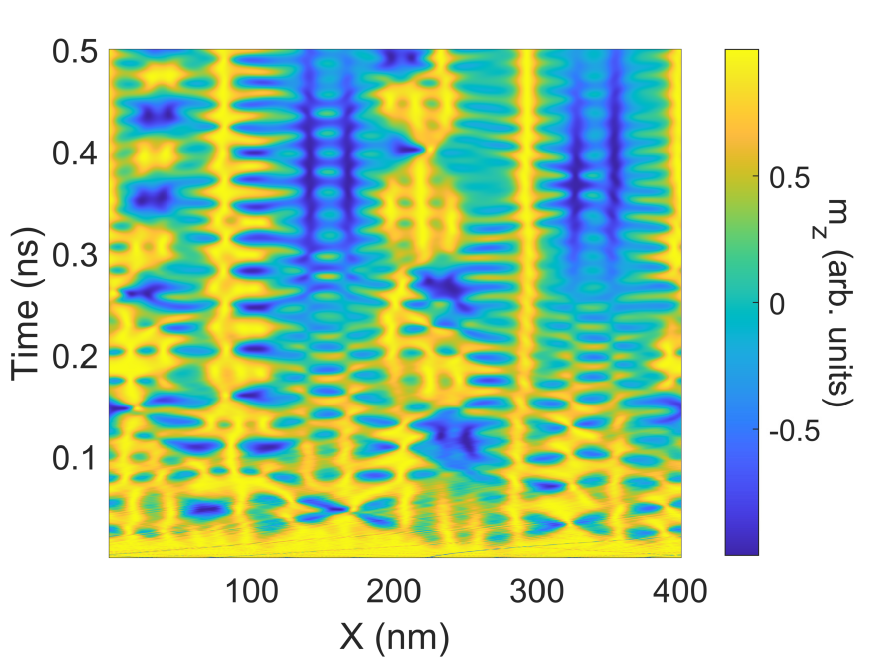}
\includegraphics[width=2.5in]{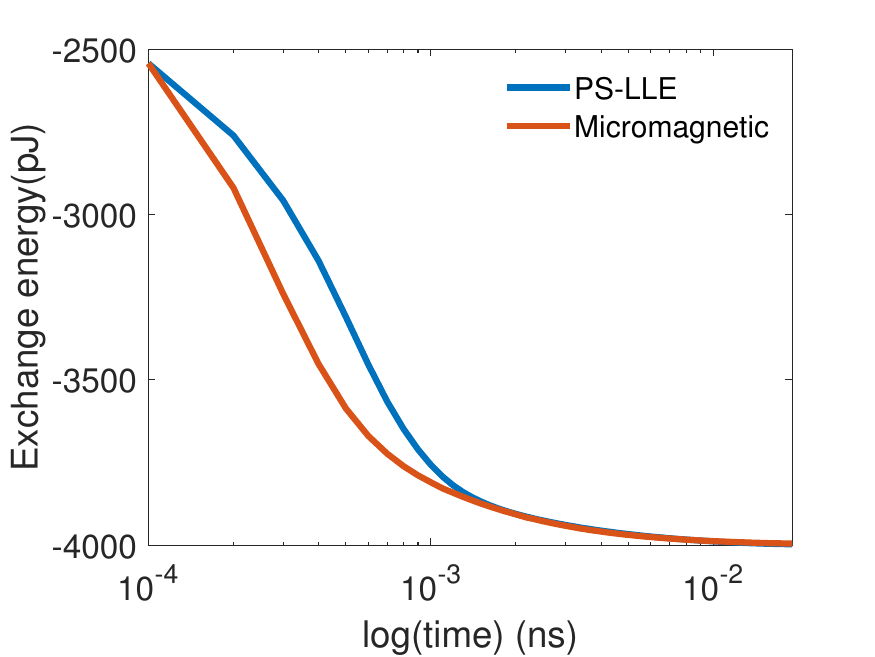}
\includegraphics[width=2.5in]{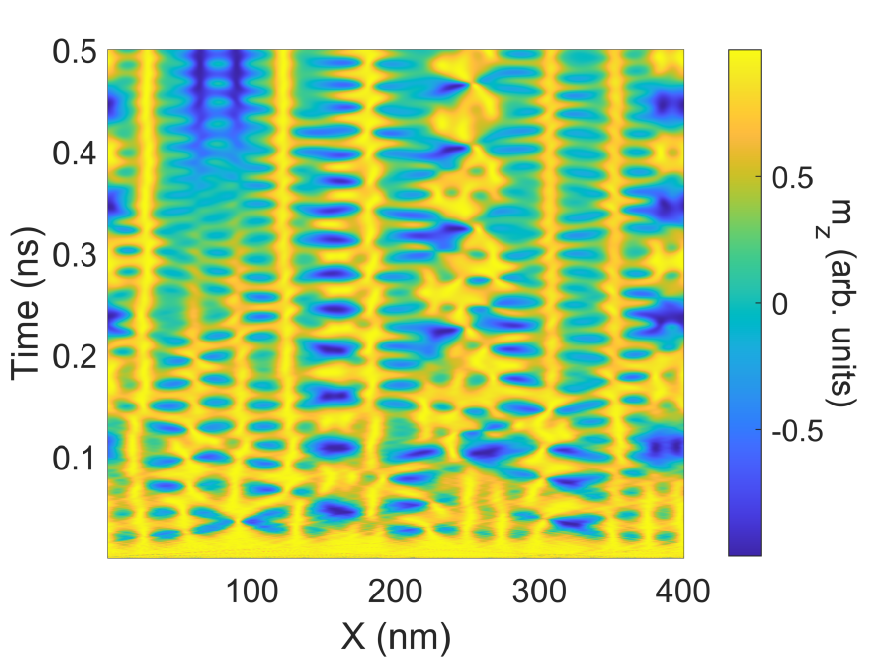}
\includegraphics[width=2.5in]{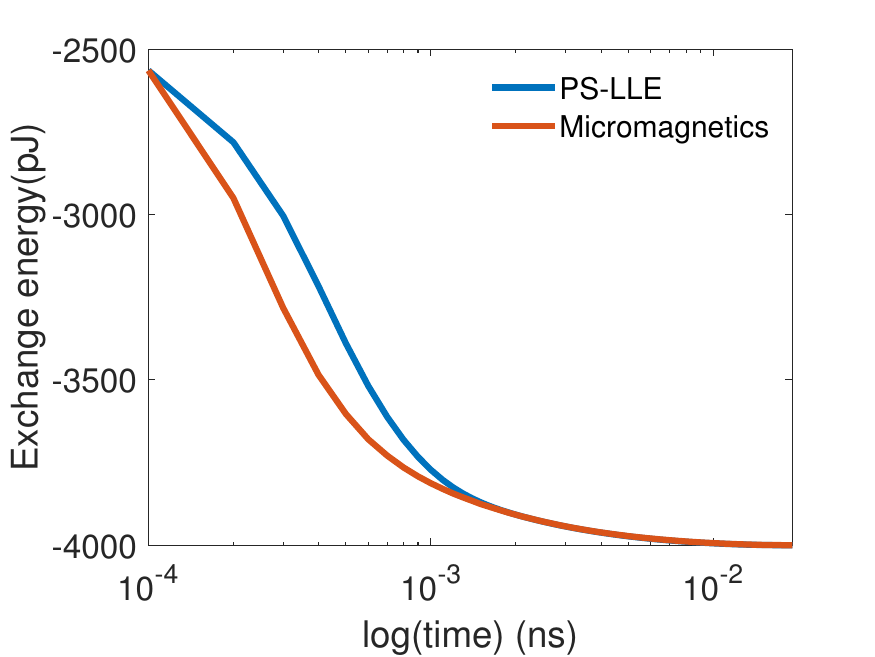}
\caption{Resulting magnetization and energy profiles from 1, 8, and 16 pulses on the same $400$~nm~$\times 1$~nm~$\times 1$~nm CoFe/Ni nanowire modeled before. However, these simulations were generated with in-plane pulses; this generated modulational instability(MI) as the time evolved.\label{fig:si_MI}}
\end{figure}
Fig.~\ref{fig:si_MI} shows the results from  simulated cases of 1, 8, and 16 pulses applied to the same $400$~nm~$\times 1$~nm~$\times 1$~nm CoFe/Ni ferromagnetic nanowire as before. These plots correspond to the evolution of the magnetization across the simulated time of $0.5$~ns. In these cases, the pulses were applied in-plane and thus the evolution over time is viewed from the $m_{x}$ perspective, where the $m_{z}$ component is represented via the colorbars. On the right side is the energy over time, computed the same as in the out-of-plane simulations from before. Unlike before, this case represents a metastable configuration for the PMA CoFe/Ni and, therefore, these in-plane pulses incur {in; this is a grammar error.} modulational instability (MI). Beyond the linear MI regime shown in the main text, the nonlinear regime results in the nucleation of solitons which, in a 1D ferromagnetic chain take the form of kinks or domain walls. 
These domain walls are presented in the plots on the left-side. As seen before, the discrepancy in magnon group velocities between the micromagnetic approximation (LLG) and PS-LLE can be seen in the rate at which the energy damps over the simulation time. This damping rate presents itself via the shortening of the expected energy between the two methods, with the 1 pulse case damping the slowest, and the 16 pulse profile damping the fastest. 

\section{Transient grating with temperature}
\label{sec:si_TMM}

To model a more realistic transient grating, we implement a laser profile in ASD and use the two-temperature model (TTM) to obtain the magnetization profile. For such a simulation, the following table of parameters are used:

\begin{center}
\begin{tabular}{|l|c|c|c|}
\hline Physical Constant & Symbol & Value Used & Units \\
\hline \hline Electron specific heat coefficient $\left(\gamma_e\right)$ & $C_e=\gamma_e T_e$ & 175 & $\frac{\mathrm{J}}{\mathrm{m}^3 \mathrm{~K}^2}$ \\
\hline Phonon specific heat & $C_p$ & $10^7$ & $\frac{\mathrm{J}}{\mathrm{m}^3 \mathrm{~K}}$ \\
\hline Electron-phonon coupling & $G_{ep}$ & $10^{18}$ & $\frac{\mathrm{J}}{\mathrm{m}^3 \mathrm{~K}}$ \\
\hline Laser pump fluence prefactor & $P_0$ & $10.0 \times 10^{21}$ & $\frac{\mathrm{J}}{\mathrm{m}^3 \mathrm{~s}}$ \\
\hline Laser pump temporal offset & $t_0$ & 10 & $\mathrm{ps}$ \\
\hline Laser pump temporal width & $\tau$ & 0.5 & $\mathrm{ps}$ \\
\hline Laser pump spacial width & $\sigma$ & 1.775 & $\mathrm{nm}$ \\
\hline
\end{tabular}
\end{center}

\begin{figure}[t]
\centering 
\includegraphics[width=3.6in]{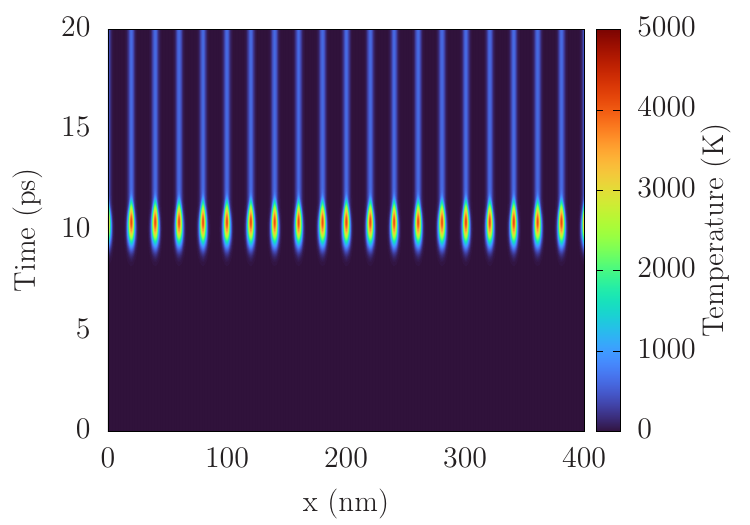}
\label{fig:ttm}
\caption{The spin chain temperature in the ASD model during the first 20~ps following laser heating with a periodicity of 20~nm.}
\end{figure}

The resulting temperature profile along the 1D chain as a function of time is shown in Fig.~\ref{fig:ttm}, where the laser peak energy arrives at $10$~ps.

\begin{figure}[t]
\centering 
\includegraphics[width=3.6in]{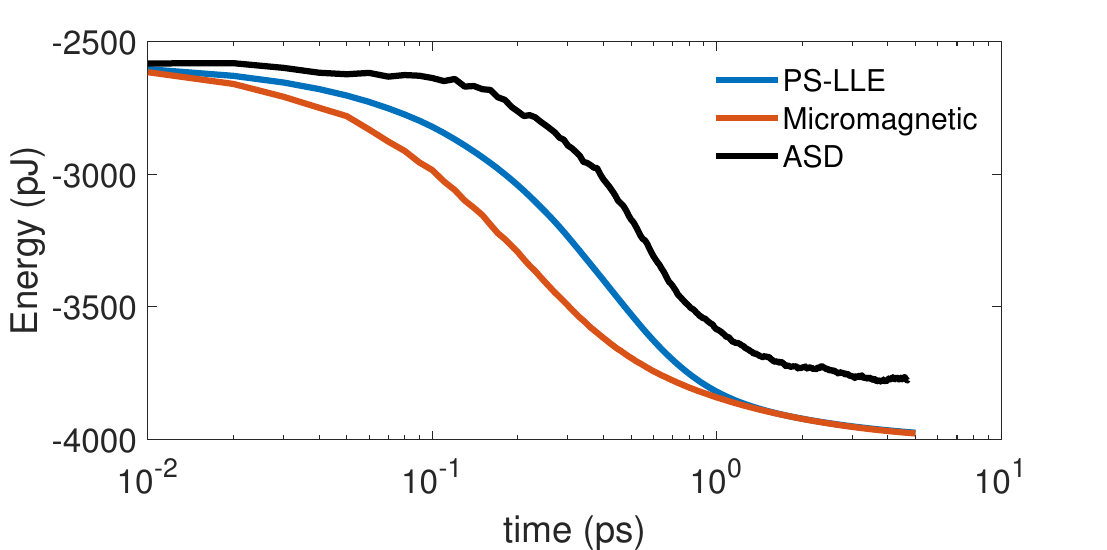}
\label{fig:EnergyTemp}
\caption{Comparison of exchange energy between PS-LLE in solid blue curve, micromagnetic approximation (LLG) in solid red curve, and ASD in solid black curve for a case where the input conditions are taken from ASD and with spin temperature implemented via the TTM.}
\end{figure}

We retrieve the magnetization profile at its maximum quench at $10.53$~ps. The exchange energy as a function of time in shown in Fig.~\ref{fig:EnergyTemp}. In this case, the PS-LLE model does not exactly agree with ASD. This is because temperature is not yet implemented in the PS-LLE model, so that the lingering temperature in the ASD model continues to randomize the spins leading to a longer transient of high exchange energy. However, the qualitative shape of the energy evolution in PS-LLE closely follows ASD upon an exponential shift. Again, the micromagnetic implementation dissipates energy more quickly for large-$k$ magnons.


\end{document}